\documentclass[amsfonts,amsmath,amssymb,showpacs]{revtex4}

\usepackage{graphicx}
\usepackage{bm}

\def\l{\lambda}
\def\ds{\bar d}
\def\dc{{{\bar d}_{\rm c}}}
\def\zm{\bar z}
\def\Pc{P}
\def\Pn{P^\circ}
\def\Pt{P^\bullet}
\def\sch{\!\!\sum_{\substack{y\text{ children} \\ \text{of }x}}}
\def\F{{\cal F}}
\def\G{{\cal G}}
\def\S{{\cal S}}
\def\T{{\cal T}}
\newcommand{\avgc}[2]{\big\langle #1 \big\rangle_{#2}}
\newcommand{\avgn}[2]{\big\langle #1 \big\rangle^\circ_{#2}}
\newcommand{\avgt}[2]{\big\langle #1 \big\rangle^\bullet_{#2}}
\newcommand{\avg}[1]{\langle #1 \rangle}

\newcommand{\bds}[1]{\boldsymbol{#1}}
\def\l{\lambda}
\def\om{\omega}
\def\Om{\Omega}

\begin{document}

\preprint{Bicocca-FT-02-12}
\title{The spectral dimension of random trees} 
\author{Claudio Destri} \email{Claudio.Destri@mib.infn.it}
\author{Luca Donetti} \email{Luca.Donetti@mib.infn.it} 
\affiliation{Dipartimento di Fisica
G. Occhialini, Universit\`a di Milano--Bicocca and INFN, Sezione di
Milano, Piazza delle Scienze 3 - I-20126 Milano, Italy}

\begin{abstract}
  We present a simple yet rigorous approach to the determination of the
  spectral dimension of random trees, based on the study of the massless
  limit of the Gaussian model on such trees. As a byproduct, we obtain
  evidence in favor of a new scaling hypothesis for the Gaussian model on
  generic bounded graphs and in favor of a previously conjectured exact
  relation between spectral and connectivity dimensions on more general
  tree--like structures.
\end{abstract}


\maketitle

\section{Introduction and summary}

The {\em spectral dimension} $\ds$ was first introduced by Alexander and
Orbach \cite{alexorb} to characterize the low frequency vibration spectrum of
fractals.  Nowadays, it is considered by most the right generalization of
the euclidean dimension of regular lattices to irregular structures in
general, whether fractal or not, such as polymers, glasses, percolation
clusters, dendritic growths and so on. This is verified at least in many
experimental situations, as well as in some theoretical model examples,
whenever the physical dimensionality explicitly enter observables such as
the spectrum of density fluctuations, the long time properties of random
walks, the critical exponents of statistical models and many others.

In a theoretical setting, we may consider a connected graph as model of
generic irregular structure, with physical microscopic degrees of freedom
attached to the nodes and interaction among them associated to the
links. This scheme implies that the coordination of nodes is bounded (the
number of nearest neighbors of an atom, molecule or basic building block
has a geometrical upper bound) and, if the graph in question is infinite,
that the surface of any Van Hove ball (defined in terms of the chemical
distance alone) must be negligible with respect to the volume when its
radius goes to infinity. In other, more physically immediate words, we say
that the structure is embeddable as a whole in a finite--dimensional
euclidean space and a standard thermodynamic limit of infinite systems
exists. We are thus led to consider {\em bounded graphs} \cite{ourselves}
and to study their spectral dimension.

To be precise, we refer here to the {\em average} spectral dimension,
that is a global property of the graph, related for instance to the
infrared singularity of the Gaussian process or, equivalently, to the
density of small eigenvalues of the Laplacian or the graph average of
the long--time tail of the random walk return probability
\cite{avgds}. On macroscopically inhomogeneous structures this average
dimension may differ from the local one, which enters the long--time
tail of the random walk return probability on a given node
\cite{avgds}. However, it is the average spectral dimension that plays
the same role of the lattice euclidean dimension in many contexts, as
for instance in providing a consistent criterion on whether a
continuous symmetry breaks down at low temperatures \cite{dstheor}.

In this work we address the problem of determining the average spectral
dimension of bounded trees and in particular of statistically homogeneous
random trees \cite{ourselves} (this homogeneity actually implies that local
and average spectral dimensions coincide), which are defined by the
following branching algorithm: given a set of coordination fractions
$\{f_z, \;1\le z \le \zm\}$ that are properly normalized and have average
coordination 2
\begin{equation} \label{eq:fnorm}
  \sum_{z=1}^{\zm} f_z = 1 \quad,\qquad  \sum_{z=1}^{\zm} z f_z = 2
\end{equation}
an infinite random tree is built by first selecting with probability $f_z$
the coordination of the origin and then, proceeding shell by shell away
from the origin, by choosing the coordination of every new node with
probability $f_z$ except for the first node of each shell, when the
modified probability $\tilde f_z = (z-1) f_z$ is used to implement the
non--extinction preconditioning (this is algorithm {\bf B} of
\cite{ourselves}). These random trees represent a generalization of the
incipient infinite percolation clusters on Bethe lattices \cite{bethe},
since the branching probabilities are freely assigned rather than being
fixed by a single percolation probability. They also generalize to
arbitrary coordination fractions $f_z$ the infinitely large specimens of
the branched polymers studied for instance in \cite{jonsson}.

Since the introduction of spectral dimension, the value of $\ds = 4/3$
has been conjectured for the incipient infinite percolation cluster in
any dimension. This conjecture was proved false at low dimensions,
while has been considered valid in the limit of infinite dimension
(percolation on Bethe lattices). Leyvraz and Stanley \cite{levstan} had
in fact produced an heuristic scaling argument in favor of $\ds = 4/3$
already in 1983. Several years later, a more rigorous approach was
pursued in \cite{jonsson}, where the branched polymer phase of
two--dimensional quantum gravity was proved to have indeed $\ds =
4/3$. Branched polymers correspond to a grandcanonical ensemble of
trees with all possible sizes, which is generated by a branching
process like ours above but without the non--extinction
preconditioning. In this case the definition itself of spectral
dimension is subtle since such concept does not apply to finite
graphs. This is the major difficulty in the approach of
\cite{jonsson}. Another mathematically well founded approach to the
problem is given in \cite{kesten}, where the exponent $\theta$ of the
anomalous diffusion on our class of random trees is determined as
$\theta=1$ (the author reports only the shorter and simpler of two
different derivations, which turns out to be more than twenty pages
long and states that the other, more complete treatment is
``monstrously long''). It is commonly held, although never rigorously
established in general, that $\theta$ is connected to the spectral
dimension and the (average) connectivity (or intrinsic Hausdorff)
dimension $\dc$ as:
\begin{equation}\label{eq:theta}
        \frac1{2+\theta} = \frac\ds{2\,\dc}
\end{equation}
Therefore $\theta=1$ would imply $\ds = 4/3$, since $\dc=2$ for
random trees \cite{ourselves}.

In this paper we provide a new, independent and fairly rigorous derivation
of $\ds = 4/3$ based on the infrared properties of the Gaussian model on
our random trees (section \ref{scaling}). The connection of the Gaussian
model with the random walk on any graph is a well known fact that we
briefly review for completeness in section \ref{gauss}. As a byproduct, we
obtain also an argument in favor of the conjecture \eqref{eq:theta}, for
any infinite graph for which the thermodynamic limit of the Gaussian model
exists: it is natural in fact to expect the scaling form
\begin{equation}\label{eq:scaling0}
  \mu \sim \mu_0r^{\dc} f(\mu_0r^\gamma)
\end{equation}
for the effective squared mass of the Gaussian model in terms of a very
small ``bare'' squared mass $\mu_0$ and the very large radius of a graph
(see section \ref{gauss} for the proper definitions); moreover we
conjecture also that the finite--size exponent $\gamma$ coincides with
$2+\theta$, which is called sometimes {\em random walk dimension} and
characterizes the scaling $t\sim r^{2+\theta}$ between the time needed by a
random walker to visit with non--negligible frequency nodes at distance $r$
from its starting point; then the existence of the thermodynamic limit on
the effective mass implies that $f(x)\sim x^{-\dc/\gamma}$ and so
\begin{equation*}
  \mu \sim \mu_0^{1- \dc/\gamma}
\end{equation*}
yielding exactly equation \eqref{eq:theta}, since in general
$\mu\sim\mu_0^{1-\ds/2}$ as $\mu_0\to0$ (see section \ref{gauss}).

In our treatment of random trees we are indeed able to explicitly derive
the scaling law \eqref{eq:scaling0} with $\dc=2$ and $\gamma=\dc+1=3$,
demonstrating $\ds=4/3$. It is interesting to notice that if the scaling
law \eqref{eq:scaling0} is assumed a priori, with $\dc=2$ as well known for
random trees, then the precise numerical value of $\gamma$ (and therefore
of $\ds$) follows from a very simple ``geometrical'' fact: as shown in
section \ref{scaling}, the first correction to the finite tree behaviour
$\mu\propto\mu_0r^{\dc}$ comes from the sum of squared volumes of all
subtrees, which grows like $r^{2\dc+1}$ if the volume of tree grows like
$r^\dc$. Then
\begin{equation}\label{dsdc}
  \ds = \frac{2\dc}{\dc+1}
\end{equation}
which is also a known conjecture, even with $\dc\neq2$, for the incipient
percolation cluster and other tree--like disordered structures \cite{shlomo}.
Notice that $\ds$ is monotonically increasing from $1$ to $4/3$ as 
$\dc$ grows from $1$ to $2$. 

Let us observe that $\dc=2$ and $\ds=4/3$ are the greatest value among
all known average connectivity and spectral dimensions for bounded
trees. This fact lead us to formulate the following conjecture: for
all bounded trees $\dc\le2$ and $\ds \le \frac43$, with saturation
only in the case of random trees.  Evidently this must be true only
for the {\em average} dimensions; there indeed exist examples of trees
(such as $NT_D$ \cite{NTD}) whose {\em local} connectivity and
spectral dimension, $d_{\rm c}$ and $d$, exceed $2$ and $4/3$,
respectively; these present however macroscopic inhomogeneities so
that the local dimensions are different from the average ones which do
satisfy the bounds at $2$ and $4/3$. Let us observe also that equation
\eqref{dsdc} does not apply in the same examples if the average
dimensions are replaced by the local ones.

At the moment these upper bounds are only conjectured, but it has been
rigorously proved that $\ds \le 2$, since bounded trees are always
recurrent on the average \cite{roa}.

\section{Gaussian model, random walks and spectral dimension}
\label{gauss}
The Gaussian model on a generic connected infinite graph $\G$ is defined
\cite{hhw} by assigning a real--valued random variable $\phi_x$ to each
node $x\in \G$ with the following probability distribution
\begin{equation}\label{eq:gaussian}
        d\nu_r[\phi] = \frac1{Z_r} 
        \exp\Bigl[-\tfrac12\!\!\!\! \sum_{x,y \in B(o,r)}\phi_x (L_{xy} + 
        \mu_0\delta_{xy})\phi_y \Bigr] \prod_{x\in B(x,r)} d\phi_x
\end{equation}
for the collection $\phi=\{\phi_x\,;\,x\in B(o,r)\}$ relative to any
Van Hove ball
\begin{equation*} 
        B(o,r)=\{y\in \G: d(o,y)\le r \} \;,\quad o \in\G\;,\quad 
        0 \le r< \infty 
\end{equation*}
where $d(o,y)$ is the chemical distance. In equation
\eqref{eq:gaussian} $Z_r$ is the proper normalization factor (the
partition function), $\mu_0>0$ is a free parameter (the squared mass
in field--theoretic sense) and
\begin{equation}\label{eq:laplacian}
        L_{xy} = z_x \delta_{xy} - A_{xy}
\end{equation}
is (minus) the Laplacian matrix on $\G$, $z_x$ being the coordination or
degree of $x$ and $A_{xy}$ the adjacency matrix).

The thermodynamic limit is achieved by letting $r\to\infty$ and defines a
Gaussian measure for the whole field $\phi=\{\phi_x\,;\,x\in\G\}$ which
does not depend on the center $o$ of the ball \cite{rimtim} if $\G$ is
bounded. The covariance of this Gaussian process reads
\begin{equation}\label{eq:gauss2}
        \langle \phi_x \phi_y \rangle \equiv  C_{xy}(\mu_0) = 
        (\bds L + \mu_0)^{-1}_{xy}
\end{equation}
and setting
\begin{equation}\label{eq:C2W}
        C_{xy} = \frac{(1-\bds W)^{-1}_{xy}}{z_x + \mu_0} ,~~~
        W_{xy} = \frac{A_{xy}}{z_y + \mu_0} 
\end{equation}
one obtains the standard connection with the random walks over
$\G$ \cite{hhw}:
\begin{equation}
\label{eq:RW}
        (1-\bds W)^{-1}_{xy} \,=\,\sum_{t=0}^\infty \, (\bds W^t)_{xy} = 
        \sum_{\gamma:\,x\leftarrow y} W[\gamma]
\end{equation}
where the last sum runs over all paths from $y$ to $x$, each weighted by
the product along the path of the one--step jump probabilities in $\bds W$:
\begin{equation}
        \gamma = (x,y_{t-1},\ldots,y_2,y_1,y) \Longrightarrow
        W[\gamma] = W_{xy_{t-1}} W_{y_{t-1}y_{t-2}},\ldots,W_{y_2y_1}W_{y_1y}
\end{equation}
Notice that, as long as $\mu_0>0$, we have $\sum_x (\bds W^t)_{xy}<1$
for any $t$, namely the walker has a nonzero death probability. This
implies that $C_{xy}$ is a smooth functions of $\mu_0$ for
$\mu_0\ge\epsilon>0$. In the limit $\mu_0\to 0$ the walker never dies
and the standard random walk is recovered; then the sum over paths in
equation (\ref{eq:RW}) is dominated by the infinitely long paths which
sample the large scale structure of the entire graph (``large scale''
refers here to the metric induced by the chemical distance alone).
This typically reflects itself into a singularity of $C_{xy}$ at
$\mu_0=0$. Consider for instance a diagonal element $C_{xx}$ of the
covariance and suppose that its singularity is power--like
\begin{equation}\label{eq:leading}
        {\rm Sing}\, C_{xx}(\mu_0) \propto  \mu_0^{d/2-1} 
\end{equation} 
Then by equation \eqref{eq:C2W} one sees that the same power--like
behaviour appears, as $\l\to 1$, in the discrete Laplace transform
$P_{xx}(\l)$ of the random walks return probability $P_{xx}(t)$ at
$x$. In turn this implies the long time behaviour
\begin{equation}\label{eq:leading2}
        P_{xx}(t) \propto  t^{-d/2} \;,\quad t\to\infty
\end{equation}
identifying $d$ with the local spectral dimension, which does not depend on
the specific node $x$ \cite{hhw}. To be precise the denomination ``spectral
dimension'' refers to the behaviour of the spectral density $\rho(l)$ of
low-lying eigenvalues of the Laplacian $\bds L$, namely $\rho(l)\sim
l^{~\ds/2 -1}$. In fact, it can be shown \cite{avgds} that $\ds$ is
connected to the long time behaviour of the graph average of $P_{xx}(t)$, or
equivalently to the infrared power--like singularity of the graph average
$\overline{C(\mu_0)}$ of $C_{xx}$, that is
\begin{equation}\label{eq:leading3}
   {\rm Sing}\, \overline{C(\mu_0)} = {\rm Sing}\,\lim_{r\to\infty}
   \frac1{|B(o,r)|}\sum_{x\in B(o,r)} C_{xx}(\mu_0) 
   \propto  (\mu_0)^{{\bar d}/2-1}
\end{equation}
where $|B(o,r)|$ is the volume of $B(o,r)$. $\ds$ is therefore called
sometimes ``average'' spectral dimension. On regular lattices it coincides
with the local spectral dimension and the usual Euclidean dimension.

By definition, the diagonal elements $C_{xx}$ of the covariance may
be written as
\begin{equation}\label{eq:Zx}
  C_{xx} = \int d\varphi\, \varphi^2 \,{\cal Z}_x(\varphi)
\end{equation}
where
\begin{equation*}
  {\cal Z}_x(\varphi) =  \lim_{r\to\infty}\int d\nu_r[\phi]\,
  \delta(\phi_x-\varphi)  
\end{equation*}
Thanks to the fundamental self--reproducing property of Gaussian
integrations, ${\cal Z}_x(\varphi)$ is necessarily a normalized Gaussian in
$\varphi$, that is
\begin{equation*}
  {\cal Z}_x(\varphi) = \left(\frac{\mu_x}{2\pi}\right)^{1/2}\, 
  e^{-\mu_x\varphi^2/2}
\end{equation*}
where, for consistency with equation \eqref{eq:Zx}
\begin{equation}\label{eq:mu2C}
  \mu_x(\mu_0) = \frac1{2\,C_{xx}(\mu_0)}
\end{equation}
is the ``effective mass'' at $x$.  

\section{The Gaussian model on random trees}

Let us now consider the Gaussian model on an infinite bounded tree $\T$.
Any node $x\in\T$ may be regarded as the root of an infinite branching
process which produces $\T$ as family tree rooted at $x$. This simply means
that $x$ has as many descendants, or children, as its coordination $z_x$,
while any other node $x'$ has $z_{x'}-1$ children one step further away from
the unique ancestor $x$ and a father one step closer to $x$. In particular,
the $z_x$ children $y$ of $x$ are themselves roots of disjoint subtrees, or
branches $\T'_y$, which may or may not be infinite. Now, by the simple
rules of Gaussian integration, one easily verifies the following relation
between the effective mass at $x$, relative to the entire $\T$ and the
effective masses at children $y$ of $x$, each one relative to the
corresponding branch $\T'_y$ :
\begin{equation} \label{eq:mucomp}
  \mu_x[\T] = \mu_0\, + \sch 
  \frac{\mu_y[\T'_y]}{1+\mu_y[\T'_y]}
\end{equation}
A similar relation holds for the effective masses $\mu_y[\T'_y]$ in terms
of the $z_y-1$ masses of $y$'s children, and so on for the rest of the
tree.

Next, suppose that the branching process is the Galton--Watson random
one, with the non--extinction preconditioning, described in the
Introduction (and in more detail in \cite{ourselves}), so that $\T$ is
an infinite random tree. Because of the preconditioning, $\T$ may be
regarded as the union of infinitely many subtrees generated by
independent and identically distributed (i.i.d.) random branching
processes, without the non--extinction preconditioning, whose roots
are attached to the nodes of a half--infinite chain (the spine) that
starts from $x$. This means that only one of the branches at $x$ is a
priori guaranteed to extend to infinity, while the other $z_x-1$ are
the family trees of i.i.d. random branching processes which might stop
after finitely many generations.

Let us consider first the probability distributions relative the branching
process itself and in particular the probability $P_r(\mu)$ that the
effective mass at the root $x$ be exactly $\mu$ when the branching
algorithm is iterated for $r$ generations, that is
\begin{equation*}
  P_r(\mu) = \avg{\delta(\mu_x -\mu)}_r
\end{equation*}
where the average is over all realizations of the tree with $r$
generations, namely on all possible histories of the branching
algorithm with $r$ shells completed. Taking equation \eqref{eq:mucomp}
and the spinal decomposition into account, we may write
\begin{equation}\label{eq:P}
  P_{r+1}(\mu) = \sum_z f_z \int d\mu_1 d\mu_2 \ldots d\mu_z \,
  \delta\Bigl(\mu-\mu_0-\sum_{j=1}^z\frac{\mu_j}{1+\mu_j} \Bigr)
  \Pt_r(\mu_1) \prod_{j=2}^z \Pn_r(\mu_j) 
\end{equation}
where $\Pn(\mu)$ is the same probability without the non--extinction
preconditioning and fulfills the functional recursion
\begin{equation}\label{eq:Pn}
  \Pn_{r+1}(\mu) = \sum_z f_z \int d\mu_1 d\mu_2 \ldots d\mu_{z-1} \,
  \delta\Bigl(\mu-\mu_0-\sum_{j=1}^{z-1}\frac{\mu_j}{1+\mu_j} \Bigr)
  \prod_{j=1}^{z-1} \Pn_r(\mu_j)
\end{equation}
while $\Pt_r(\mu)$ is the probability when also the coordination of
the root is extracted with the modified probability ${\tilde f}_z$ and
satisfies therefore
\begin{equation}\label{eq:Pt}
  \Pt_{r+1}(\mu) = \sum_z {\tilde f}_z \int d\mu_1 d\mu_2 \ldots d\mu_{z-1} \,
  \delta \Bigl(\mu-\mu_0-\sum_{j=1}^{z-1}\frac{\mu_j}{1+\mu_j} \Bigr)
  \Pt_r(\mu_1) \prod_{j=2}^{z-1} \Pn_r(\mu_j) 
\end{equation}
As $r\to\infty$ we expect all these probability distributions to converge
to their limiting forms $P_\infty(\mu)$, $\Pn_\infty(\mu)$ and $\Pt
_\infty(\mu)$, turning eqs. \eqref{eq:P}, \eqref{eq:Pn} and \eqref{eq:Pt}
into highly nontrivial coupled integro--functional equations for
$P_\infty(\mu)$, $\Pn_\infty(\mu)$ and $\Pt_\infty(\mu)$ with a
parametric dependence on $\mu_0$. In particular we are interested in their
infrared behaviour as $\mu_0\to0$.

\section{Statistical homogeneity and auto--averaging property}

Before addressing the just mentioned problem, we need to establish the
precise relation with the original problem, that of determining the
$\mu_0\to0$ behaviour of $\overline{C(\mu_0)}$ on a given ``generic''
infinite random tree $\T$ produced by our algorithm. By ``generic'' in this
context we mean an infinite $\T$ belonging to the subset of unit measure of
trees that fulfill the {\em auto--averaging property} for local
observables. To be explicit, consider the frequency $F(r,\tau)$ of
occurrence in $\T$ of a given rooted tree $\tau$ with $r$ generations. In
terms of the Van Hove balls $B(x,r)$, regarded as subtrees of $\T$, we may
write $F(r,\tau)$ as the graph average
\begin{equation*}
  F(r,\tau) = \lim_{R\to\infty} F_R(r,\tau) \;,\quad 
  F_R(r,\tau) = \frac1{|B(o,R)|}\sum_{x\in B(o,R)} \delta[B(x,r)=\tau]
\end{equation*}
where $o$ is an arbitrary node of $\T$ and $\delta[\tau'=\tau]$ is one if
its argument is a true statement and zero otherwise. $F(r,\tau)$ does not
depend on the choice of $o$ ($\T$ is a bounded graph) and it is convenient
to identify it $o$ with the root, or origin, of the infinite
branching algorithm. 

Let us then recall that any $\T$ produced by the algorithm may be
decomposed into the spine $\S$ (the origin $o$ plus all the nodes on which
the coordination was extracted with ${\tilde f}_z$) and the {\em
  transverse} family trees of the identical, independent and non
preconditioned branching processes starting from the nodes of the spine
(see \cite{ourselves}\cite{kesten} for more details). Evidently, the
interesection of the transverse family trees rooted at $s\in\S$ with
$B(o,R)$ contains at most $R-d(s,o)$ generations, so that they contribute
differently, even on average, to $F_R(r,\tau)$. Therefore, to simplify our
derivation, we consider a modified thermodynamic limit, in which the
infinite tree is not recovered through a Van Hove ball with diverging
radius, but as $R\to\infty$ limit of the set
\begin{equation*}
  \Om_R = \bigcup_{s\in\S_R} \om_{s,R}
\end{equation*}
where $S_R$ is the restriction of the spine to the first $R$ generations of
$\T$ and $\om_{s,R}$ are the union of all transverse family trees rooted at
$s$ and restricted to the first $R$ generations. Thus we redefine
$F_R(r,\tau)$ as 
\begin{equation*}
  F_R(r,\tau) = \sum_{s\in\S_R} w_{s,R}\, Q_{s,R}(r,\tau) 
  \;,\quad w_{s,R} = \frac{|\om_{s,R}|}{|\Om_R|} \;,\quad 
  Q_{s,R}(r,\tau) = \frac1{|\om_{s,R}|}\sum_{x\in\om_{s,R}} \delta[B(x,r)=\tau]
\end{equation*}
Notice that, by construction, the random weights $w_{s,R}$ are all
identically distributed and are all independent except for the
normalization constraint $\sum_{s\in\S_R} w_{s,R}=1$. The random
frequencies $Q_{s,R}(r,\tau)$ are all identically distributed, but are
not independent because a ball $B(x,r)$ will in general intersect
several $\om_{s,R}$ if $d(x,s)< r$. Therefore we split $\om_{s,R}$
into $\om_{s,R}^>$, formed by nodes $x$ such that $d(x,s)> r$, and its
complement $\om_s^\leqslant$, which does not depend on $R$ for large
$R$.  This implies an analogous splitting for the random weights and
the random frequencies. Now the random frequencies $Q_{s,R}^>(r,\tau)$
are all independent since the two events $B(x,r)=\tau$ and
$B(x',r)=\tau$ are independent if $x \in \om_{s,R}^>$ and $x' \in
\om_{s',R}^>$ with $s\neq s'$. Notice also that $Q_{s,R}(r,\tau)$
and $w_{s',R}$ are independent when $s\neq s'$.

Thus we have
\begin{equation*}
  F_R(r,\tau) = F_R^\leqslant(r,\tau) + F_R^>(r,\tau)
\end{equation*}
with $F_R^\leqslant(r,\tau)$ getting contributions solely from nodes at
distance not larger than $r$ from the spine. Clearly, in the limit
$R\to\infty$, these nodes form a subset of zero measure of $\T$, since
$\bigcup_s \om_s^\leqslant(r,\tau)$ is essentially one dimensional while a
generic $\T$ has $\dc=2$. Hence, as $R\to\infty$,
\begin{equation*}
  F(r,\tau) = \lim_{R\to\infty} F_R^>(r,\tau)
\end{equation*}
By analogous dimensionality arguments we get that almost always
$\lim_{R\to\infty}w_{s,R} = 0$, although the sum of all these random
weights stays properly normalized.

We may now invoke directly the law of large numbers for $F_R^>(r,\tau)$,
which is the weighted average of i.i.d. random variables. Hence, for
any {\em finite} $r$, $F(r,\tau)$ is almost always non--fluctuating on
infinite trees and 
\begin{equation*}
  F(r,\tau) = \avgn{Q^>(r,\tau)}{}
\end{equation*}
where $Q^>(r,\tau)$ is anyone of the $\lim_{R\to\infty} Q_{s,R}^>(r,\tau)$
and the average $\avgn{\cdot}{}$ is taken with respect to the algorithm
without the preconditioning on non--extinction. But any node $x$ on such
trees laying at a distance larger then $r$ from the origin enjoys {\em
  statistical homogeneity} for the event $B(x,r)=\tau$, since the branching
is i.i.d.  on every node (but the origin, at most) and $B(x,r)=\tau$ cannot
contain the origin. Moreover, the unique path connecting $x$ to $o$ plays
the role of the spine for a branching process rooted at $x$ and
preconditioned on non--extinction for at least $r$ generations, since the
trees contributing to $\avgn{Q^>(r,\tau)}{}$ have at least $r$
generations. We conclude therefore that almost always an infinite random
tree $\T$ has the auto--averaging property
\begin{equation}
  F(r,\tau) = \avg{\delta[B(o,r)=\tau]}
\end{equation}
with respect to the probability itself that the first $r$ shells of $\T$
build up to form the given finite tree $\tau$. Then this property holds for
any local observable and in particular for the effective mass
$\mu_x[B(x,r)]$ of a Gaussian models restricted to $B(x,r)$. Thus the
probability distribution $P_r(\mu)$ coincides with the frequency of the
event $\mu_x[B(x,r)]=\mu$, as $x$ varies throughout a ``generic'' infinite
random tree $\T$. The same would evidently apply to the probability
distribution of the variable $(2\mu)^{-1}$ and the frequency of the event
$C_{xx}[B(x,r)]=(2\mu)^{-1}$.

\section{Scaling}
\label{scaling}

Let us now return to the recursion relations \eqref{eq:P}, \eqref{eq:Pn}
and \eqref{eq:Pt} for the probability distributions $P_r(\mu)$,
$\Pn_r(\mu)$ and $\Pt_r(\mu)$, respectively. We are interested
in the thermodynamic limit $r\to\infty$ of infinite trees followed by the
massless limit $\mu_0\to0$. This two limits can actually be applied
simultaneously, provided we identify the correct scaling variables.
Suppose in fact that, as $r\to\infty$ {\em and} $\mu,\,\mu_0\to0$, the
scaling law \eqref{eq:scaling0} applies; since $\mu$ is a really a 
random variable, a better formulation is in terms of its probability, that
is 
\begin{equation}\label{eq:scaling}
  P_r(\mu;\mu_0) \simeq \mu_0^\beta\, F(\mu\mu_0^\beta,\mu_0r^\gamma)
\end{equation}
for suitable scaling exponents $\beta$ and $\gamma$ and scaling
function $F(u,s)$. $\beta$ is then directly related to the spectral
dimension, as $\beta = -1+\ds/2$, thanks to equation
\eqref{eq:leading}, the results of the previous section and the
existence of the thermodynamic limit; $\gamma$ provides instead a
measure of finite--size effects at very large sizes. In particular, if
we let $\mu_0\to0$ too fast, so that $\mu_0r^\gamma\to0$ even if
$r\to\infty$, we must recover the massless behaviour characteristic of
finite trees. This amounts to linearize the effective mass formula
\eqref{eq:mucomp}, so that
\begin{equation*}
  \mu_x[\T] = \mu_0 v[\T]
\end{equation*}
where $v[\T]$ is the volume of $\T$. Thus 
\begin{equation*}
  P_r(\mu;\mu_0) \simeq \mu_0^{-1} {\hat P}_r(\mu/\mu_0) = 
  \mu_0^{-1} {\hat P}_r(v)
\end{equation*}
where ${\hat P}_r(v)$ is the volume probability at radius $r$, which for
large $r$ has a scaling form in terms of $v/r^2$ explicitly determined in
\cite{ourselves} (thus implying $\dc=2$). Since this scaling form is
highly nontrivial, with exponentially small behaviour also for $v/r^2\to0$,
compatibility with the scaling hypothesis \eqref{eq:scaling} requires
$1+\beta = 2/\gamma$, that is
\begin{equation*}
  \frac1\gamma = \frac\ds4 = \frac\ds{2\,\dc}
\end{equation*}
One is therefore lead to the identification (compare with equation
\eqref{eq:theta} in the introduction)
\begin{equation*}
  \gamma = 2+ \theta
\end{equation*}
relating finite--size effects to the exponent of anomalous diffusion. Of
course, since the relation \eqref{eq:theta} is only a well founded
conjecture, an explicit derivation of $\gamma = 2+ \theta$ would provide a
proof of the conjecture for our random trees. Let us stress the point:
$\gamma$ measures the scaling behaviour of the Gaussian model (and therefore
of the random walk) on finite random trees of very large radius $r$; $2+
\theta$ characterizes instead the scaling $t\sim r^{2+\theta}$ between the
time needed by a random walker to visit with non--negligible frequency nodes
at distance $r$ from its starting point. The two concepts are certainly
related but quite distinct, so that the identification $\gamma = 2+ \theta$
has a nontrivial content.

Let us come back now to the opposite limiting behaviour of the probability
distributions that is the subject of this work, namely $r\to\infty$ and
$\mu,\,\mu_0\to0$ with $u=\mu\mu_0^\beta$ fixed and $s=\mu_0r^\gamma$
finite and possibly very large. We shall show now that the scaling law
\eqref{eq:scaling} is indeed correct with the precise values
\begin{equation}\label{eq:result}
  \beta = -1/3 \;,\quad \gamma = 3
\end{equation}
implying $\ds=4/3$. The detailed proof, by induction on all moments of
$P_r(\mu)$, $\Pn_r(\mu)$ and $\Pt_r(\mu)$ is rather simple albeit
notationally a bit cumbersome and is reported in the Appendix. Here we only
provide a sketch of the derivation.

The first step consists in rewriting the recursion relations \eqref{eq:P},
\eqref{eq:Pn} and \eqref{eq:Pt} into a more tractable form. In fact
they are amenable to a direct treatment via Laplace transform only in the
limit $\mu_0\to0$ at fixed $r$, while we are interested in a regime where
the nonlinearity in the effective mass formula \eqref{eq:mucomp} plays a
crucial role. Therefore we consider the expansion in all powers of $\mu_0$ 
\begin{equation} \label{eq:muexp}
  \mu_x[\T] = \sum_{n=1}^{\infty} (-1)^{n+1} V_n(x) \mu_0^n
\end{equation}
where the coefficients $V_n(x)$ are integer--valued random variables
satisfying composition laws obtained by equating powers of $\mu_0$ on both
sides of equation \eqref{eq:mucomp}. These read
\begin{equation*}
\begin{split}
  V_1(x) & = 1 + \sch V_1(y) \\
  V_2(x) & = \sch \left[V_2(y) + V_1(y)^2\right] \\
  V_3(x) & = \sch \left[V_3(y) + 2 V_1(y) V_2(y) + V_1(y)^3\right] \\
  V_4(x) & = \sch \left[V_4(y) + 2 V_1(y) V_3(y) + V_2(y)^2 + 3 V_1(y)^2 V_2(y) +
   V_1(y)^4\right] \\
  V_5(x) & = \ldots
\end{split}
\end{equation*}
which we write in general as 
\begin{equation} \label{eq:fdef}
  V_n(x) = \delta_{n,1} + \sch \F_n(V_1(y),V_2(y),\ldots,V_n(y))
\end{equation}
where the form of $\F_n$ can be easily induced from the previous formulas.
Notice that $V_1$ can be interpreted as the volume of the subtree, $V_2$ as
the sum of the squares of all sub-volumes and so on.  Now we may write
recursion relations for the probability of the coefficients $V_n(x)$
themselves. For instance, in the branching algorithm without
non--extinction preconditioning, we have
\begin{equation}\label{eq:PPn}
  \Pn_{r+1}(V_1,\ldots,V_n) = \sum_z f_z 
  \!\!\!\sum_{\substack{V_{1,1} \cdots V_{1,z-1} \\ 
      V_{2,1} \cdots V_{2,z-1} \\
      \cdots \\
      V_{n,1} \cdots V_{n,z-1}}}
  \prod_{i=1}^{z-1} \Pn_r(V_{1,i},\ldots,V_{n,i})
  \prod_{k=1}^n \delta\bigl(
  V_k - \delta_{n,1} - \sum_{j=1}^{z-1} \F_k(V_{1,j},\ldots,V_{k,j})\bigr) 
\end{equation}
with similar expressions for probabilities $P_r(V_1,\ldots,V_n)$
and $\Pt_r(V_1,\ldots,V_n)$ corresponding to those for the
effective mass $P_r(\mu)$ and $\Pt_r(\mu)$. For brevity we do not
write them here; they can be found in the Appendix.

The advantage of this new formulation stands in the fact that
$\F_n(\{V_{1,i},\ldots,V_{n,i})$ is by construction linear in
$V_{n,i}$.  This allows to set up a systematic control of the large
$r$ behaviour of the recursion relations for the multiple moments of
$P_r(V_1,\ldots,V_n)$, $\Pn_r(V_1,\ldots,V_n)$ and
$\Pt_r(V_1,\ldots,V_n)$.  Consider for example the lowest moments:
from equation \eqref{eq:PPn} and its analogs in the Appendix we get
\begin{equation*}
\begin{split}
  \avgn{V_1}{r+1} & = \sum_z f_z \left( 1 + 
    (z-1)\avgn{V_1}{r}\right) = 1 + \avgn{V_1}{r} \\
  \avgt{V_1}{r+1} & = \sum_z (z-1)f_z \left(1+ \avgt{V_1}{r} +
    (z-2)\avgn{V_1}{r}\right)
  = 1 + \avgt{V_1}{r} + g''(1) \avgn{V_1}{r} \\    
  \avgc{V_1}{r+1} & = \sum_z f_z \left(1+ \avgt{V_1}{r} +
    (z-1)\avgn{V_1}{r}\right)
   = 1 + \avgt{V_1}{r} + \avgn{V_1}{r}
\end{split}
\end{equation*}
where $g(\l)$ is the so--called {\em probability generating function}
\begin{equation} \label{eq:gdef}
   g(\l)  = \sum_z  f_z \l^{z-1} = f_1 + f_2 \l +  f_3\l^2 + \ldots
\end{equation}
which thanks to eqs.~\eqref{eq:fnorm} enjoys the
properties
\begin{equation} \label{eq:gprop}
  g(1)=1 \; , \quad g'(1)=1 
\end{equation}
Hence we obtain the asymptotic behaviour for large $r$ 
\begin{equation} \label{eq:v1n}
  \avgn{V_1}{r} \sim r \;,\quad \avgc{V_1}{r} \sim \avgt{V_1}{r}\sim r^2
\end{equation}
which implies $\dc=2$ since $V_1$ is just the volume. Next consider
the averages of $V_2$ and $V_1^2$; the recursion rules read
\begin{equation*}
\begin{split}
  \avgn{V_2}{r+1} & = \sum_z (z-1) f_z  \left( \avgn{V_2}{r} +
    \avgn{V_1^2}{r} \right) = \avgn{V_2}{r} + \avgn{V_1^2}{r} \\
  \avgt{V_2}{r+1} & = \sum_z (z-1) f_z \left[ \avgt{V_2}{r} 
    + \avgt{V_1^2}{r} + (z-2) \avgn{V_2}{r} + (z-2) \avgn{V_1^2}{r} \right] \\
  & =\avgt{V_2}{r} + \avgt{V_1^2}{r} + 
  g''(1) \avgn{V_2}{r} + g''(1) \avgn{V_1^2}{r} \\
  \avgc{V_2}{r+1} & =\avgt{V_2}{r} + \avgt{V_1^2}{r} + 
  \avgn{V_2}{r} + \avgn{V_1^2}{r}\end{split}
\end{equation*}
and 
\begin{equation*}
\begin{split}
  \avgn{V_1^2}{r+1} & =  1 + \avgn{V_1^2}{r} + g''(1) 
  \left(\avgn{V_1}{r}\right)^2 + 2 \avgn{V_1}{r} \\ 
  \avgt{V_1^2}{r+1} & = 1 + \avgt{V_1^2}{r} + 2 \avgt{V_1}{r} +
  g''(1)\left[ \avgn{V_1^2}{r} + \avgn{V_1}{r} \avgt{V_1}{r} +
  2\avgn{V_1}{r} \right] + g'''(1) (\avgn{V_1}{r})^2  \\
  \avgc{V_1^2}{r+1} & = 1 + \avgt{V_1^2}{r} + 2 \avgt{V_1}{r} +
  \avgn{V_1^2}{r} + \avgn{V_1}{r} \avgt{V_1}{r} +
  2\avgn{V_1}{r}  + g''(1) (\avgn{V_1}{r})^2 
\end{split}
\end{equation*}
We can read right away the $r\to\infty$ asymptotic behaviour
\begin{align}\label{eq:v2n}
  \avgn{V_1^2}{r} & \sim r^3 & \avgn{V_2}{r} & \sim r^4  \\
  \avgc{V_1^2}{r}  & \sim r^4 & 
  \avgc{V_2}{r}  & \sim r^5 
\end{align}
Evidently the moments of $\Pt_r(V_1,\ldots,V_n)$ have all the same
asymptotic behaviour of those of $P_r(V_1,\ldots,V_n)$ and need not be
written out explicitly.

Let us make use of the results \eqref{eq:v1n} and \eqref{eq:v2n} for the
mean value of the effective mass at the root $o$:
\begin{equation}\label{eq:first2}
  \avgc{\mu}{r} = \sum_{n=1}^{\infty} (-1)^{n+1} \avgc{V_n}{r}\, \mu_0^n
  \simeq \mu_0r^2 (c_0 + c_1\mu_0 r^3 + \ldots) 
\end{equation}
where $c_0$ and $c_1$ are constants. If we {\em assume} the scaling law 
\eqref{eq:scaling} for the probability $P_r(\mu;\mu_0)$, then clearly
\begin{equation*}
  \avgc{\mu}{r} \simeq \mu_o^{-\beta} F_1(\mu_0 r^\gamma) 
\end{equation*}
with $F_1(s)$ finite in the limit $s\to\infty$; consistency with 
\eqref{eq:first2} now requires that $F_1(s)\sim s^{2/\gamma}(1+O(s^3))$ 
as $s\to 0$ to reproduce the correct $r$ behaviour on large but finite
trees; but matching also the powers of $\mu_0$ implies
\begin{equation*}
  1+\beta = 2/\gamma \;,\quad 2+\beta = 5/\gamma
\end{equation*}
yielding exactly the desired result \eqref{eq:result}. This argument 
may be turned around to avoid referring to the scaling hypothesis 
\eqref{eq:scaling} for the probability and to better clarify the role of the
thermodynamic limit. In fact, the first two terms of the $\mu_0-$expansion  
in \eqref{eq:first2} suggest that, as $r\to\infty$ and $\mu_0\to0$ at fixed
$\mu_0r^3$, 
\begin{equation}\label{eq:scaling2}
  \avgc{\mu}{r}   \simeq \mu_0r^2 G_1(\mu_0 r^3) 
\end{equation}
for a suitable scaling function $G_1(s)$. Now it is the existence of the
thermodynamic limit \cite{hhw} requires that $G_1(s)\sim s^{-2/3}$ for large
$s$, yielding again $\avgc{\mu}{\infty}\sim \mu_0^{1/3}$, that it
$\beta=-1/3$. Of course this formulation is only tentative, since we still
need to verify that, for any other $n>2$, the $n-$th term in the
$\mu_0-$expansion of $\avgc{\mu}{r}$ contains $r^{3n+2}$ as highest power
of $r$. Moreover, even if we can prove the scaling \eqref{eq:scaling2} of
$\avgc{\mu}{r}$, this does not imply the scaling \eqref{eq:scaling} for the 
whole probability $P_r(\mu;\mu_0)$. In the Appendix we demonstrate by
induction that, as $r\to\infty$ 
\begin{equation}\label{eq:basic}
    \avgc{\prod_{j=1}^{k} V_{n_j}}{r} \sim r^{3 \sum_j n_j - k} [1+O(r^{-1})]
\end{equation}
Hence, by retaining only the dominant $r-$behaviour, we have
\begin{equation*}
  \begin{split}
    \avgc{\mu}{r} & \simeq c_0 \mu_0 r^2 + c_1 \mu_0^2 r^5 + c_2 \mu_0^3 r^8 +
    \ldots + c_i \mu_0^{i+1} r^{3i+2} + \ldots \\
    & \simeq \mu_0 r^2 G_1(\mu_0 r^3)
  \end{split}
\end{equation*}
and, by the same token
\begin{equation*}
  \avgc{\mu^2}{r} = \sum_{k=2}^{\infty} (-1)^k \mu_0^k 
  \sum_{j=1}^{k-1} \avgc{V_j V_{k-j}}{r} \simeq \mu_0^2 r^4 G_2(\mu r^3)
\end{equation*}
Similarly one obtain for any $n$
\begin{equation*}
  \avgc{\mu^n}{r} \simeq \mu_0^n r^{2 n}  G_n(\mu_0 r^3)
\end{equation*}
so that, requiring the existence of thermodynamic limit, one obtains
as $\mu_0\to 0^+$ at $r=\infty$
\begin{equation*}
  \avgc{\mu^n}{} \equiv \lim_{r\to\infty} \avgc{\mu^n}{r} = b_n\mu_0^{n/3}
\end{equation*}
where
\begin{equation*}
  b_n=\lim_{s\to\infty} s^{2n/3}G_n(s)  
\end{equation*}
We may also reformulate this result as
\begin{equation*}
  \avgc{\mu^n}{r} \simeq \mu_0^{n/3} F_n(\mu_0 r^3)
\end{equation*}
where $F_n(s)=s^{2n/3}G_n(s)$ has a finite nonzero limit as $s\to\infty$.
This general form of the moments of $P_r(\mu;\mu_0)$ implies the scaling
form \eqref{eq:scaling} with $\beta = -1/3$, $\gamma = 3$ and
\begin{equation*}
  F_n(s) = \int du\, u^n F(u,s)
\end{equation*}
for the probability $P_r(\mu;\mu_0)$ itself.

Let us close this section with one more observation concerning the
universality of the scaling law \eqref{eq:scaling0} or \eqref{eq:scaling}
with respect to the coordination distribution defined by the $f_z$ numbers
or, equivalently, by the probability generating function $g(\l)$.
A more careful analysis of the dominant large $r$ terms in the recursion
rules for $V_n$ expectations shows that (see Appendix for details) 
\begin{equation*}
  \avgc{\prod_{j=1}^{k} V_{n_j}}{r} \sim  \alpha^{\sum_j n_j} \, 
  r^{3 \sum_j n_j - k} 
\end{equation*}
where $\alpha=g''(1)$ measures the fluctuations of the coordinations
since
\begin{equation*}
  \alpha = \sum_z(z-2)^2f_z
\end{equation*}
We may therefore refine the scaling law \eqref{eq:scaling} to
\begin{equation}\label{eq:scaling4}
  P_r(\mu;\mu_0) \simeq (\alpha\mu_0)^{-1/3}\, 
  F\bigl((\alpha\mu_0)^{-1/3}\mu\,,\alpha\mu_0r^3\bigr)
\end{equation}
where $F(u,s)$ is a universal scaling function.

\section{Numerical checks}

\begin{figure}
\includegraphics[height=130mm]{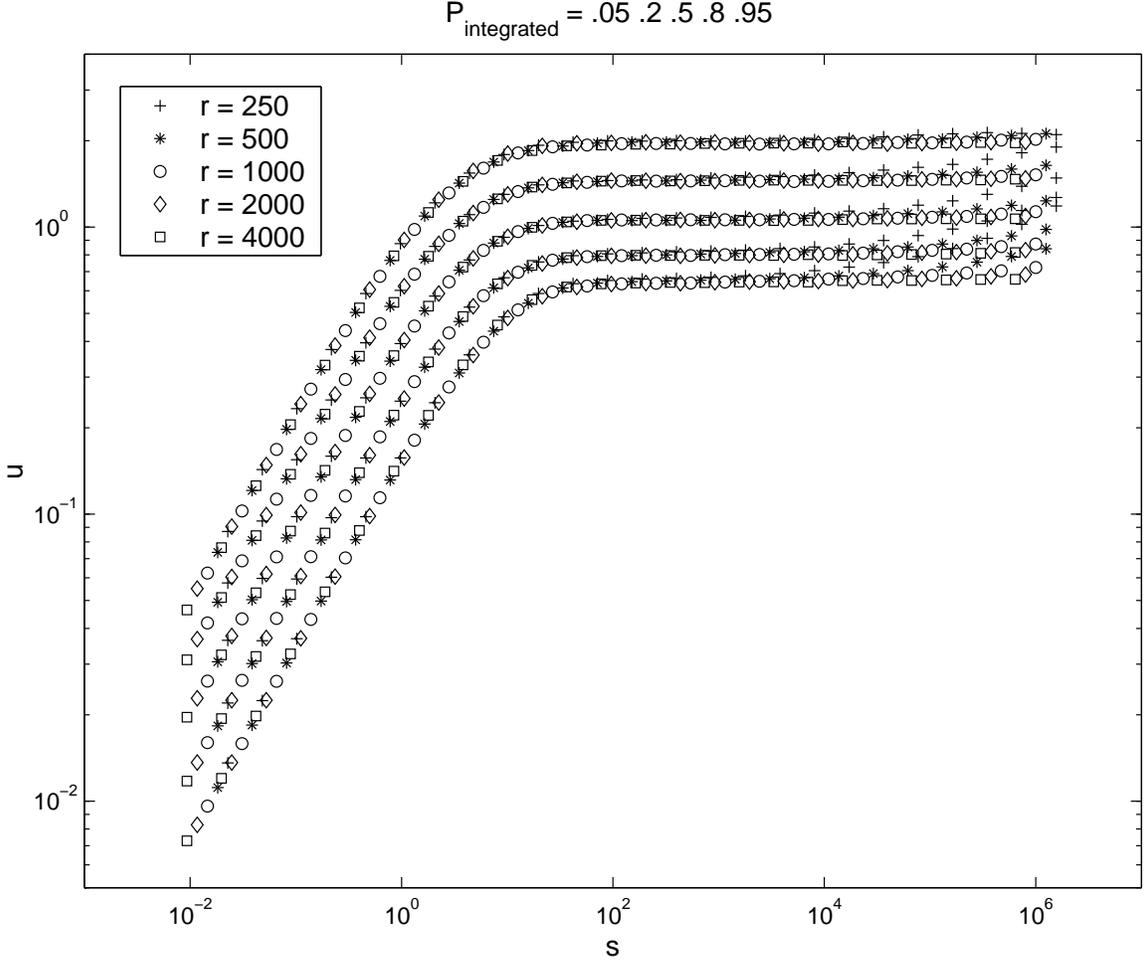}
\caption{\label{probmu} Lines of constant integrated probability from
numerical data.}
\end{figure}

\begin{figure}
\includegraphics[height=130mm]{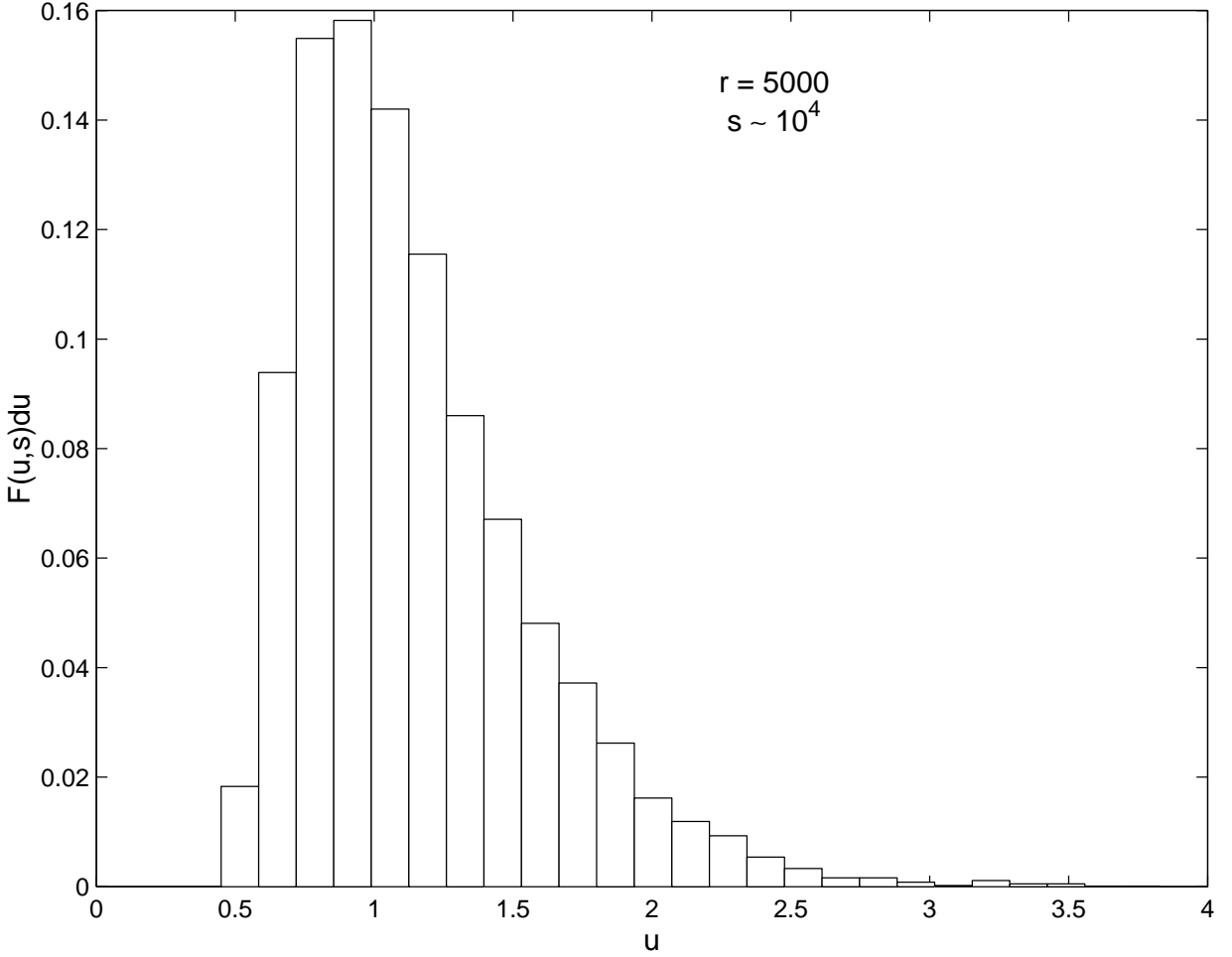}
\caption{\label{distmu} Numerical distribution for F(u,s) for 10000 
  random trees.}
\end{figure}

As an extra check on our derivation and to determine the profile of
the universal function $F(u,s)$, we numerically evaluated
$P_r(\mu;\mu_0)$ by directly calculating the effective mass $\mu$ for
each member of a large set of trees produced by our branching
algorithm. For each one of the selected values of $\mu_0$ and $r$ we
obtained several thousand values of $\mu$ that may be arranged into an
histogram providing a discretized estimate of $P_r(\mu;\mu_0)d\mu$.
This procedure does not require much computer time or memory thanks to
the recursive structure of the mass composition formula
\eqref{eq:mucomp}. Our results for the integrated probability at the
five values $0.05$, $0.2$, $0.5$, $0.8$ and $0.95$ are plotted in
figure~\ref{probmu} using the scaling variables $u=\mu_0^{-1/3}\mu$
and $s=\mu_0r^3$ (with our specific choice of $f_z$ we have
$\alpha=1$). The data show a very good scaling profile with a very
well defined crossover, as $s$ gets larger, from the behaviour typical
of small trees, $\mu\propto\mu_0$, to that characteristic of large
trees, that is $\mu\propto\mu_0^{1/3}$. The vertical section of data
at $s\simeq 10^4$ provides a good estimate of the probability
distribution $F(u,\infty)du$ and is reported in figure~\ref{distmu}.

\section{Conclusions and outlook}

Let us remark once more on the two distinct levels of our approach: our
proof by induction of equation \eqref{eq:basic} provides a rigorous basis for
the scaling form \eqref{eq:scaling} of $P_r(\mu;\mu_0)$ with $\beta = -1/3$
and $\gamma = 3$. Thanks to the auto--averaging property, this implies in
turn that indeed
\begin{equation*}
        {\rm Sing}\, \overline{C(\mu_0)} \propto  
        \mu_0^{\ds/2-1} \;,\quad \ds = 4/3
\end{equation*}
Moreover, due to the statistical homogeneity of our random trees, we see
that the local and average spectral dimension coincide. On the other hand,
if we assume scaling, as natural and common in this contexts, then the
determination $\ds = 4/3$ follows solely and very simply from the large $r$
asymptotic behaviour of $\avgc{V_2}{r}$, that is the average of the sum of
squared volumes of all subtrees of the tree. Since the volume $V_1$ scales
as $r^{\dc}$, with $\dc=2$, then $V_2$ must scale, almost by
definition, as $r^{2\dc+1}$; then
\begin{equation*}
  \avgc{\mu}{r}  \simeq \mu_0r^\dc (c_0 + c_1\mu_0 r^{\dc+1} + \ldots) 
\end{equation*} 
which is consistent with scaling only if
\begin{equation}\label{eq:general}
  \ds = \frac{2\dc}{\dc+1}
\end{equation}
Notice that this holds true even if $\dc<2$, as could happen for different
types of bounded trees. We conjectured in \cite{ourselves} that $\dc\le 2$
for any bounded tree, with saturation only for random trees.  Then
automatically $\ds\le 4/3$ for any bounded tree, with saturation only for
random trees. The general expression \eqref{eq:general} was conjectured
years ago from very different considerations \cite{shlomo}.

Another relevant issue concern the {\em spectral weight} $\bar w$ of
infinite random trees. This is the coefficient of the $\mu_0^{\ds/2-1}$
singularity of $\overline{C(\mu_0)}$ as $\mu_0\to0$, that is
\begin{equation*}
        \overline{C(\mu_0)} \sim  {\bar w} \mu_0^{-1/3}
\end{equation*}
It is clear from equation \eqref{eq:mu2C} and the auto--averaging
property that the spectral weight is related to the scaling function
$F(u,s)$ by
\begin{equation*}
  {\bar w} = \int \frac{du}{2u}F(u,\infty) 
\end{equation*}
On the other hand it is known \cite{hhw} that on any given infinite
graph with local spectral dimension $d<2$, the local spectral weight 
does not really depend on $x$, that is 
\begin{equation}
        C_{xx}(\mu_0) \sim w \mu_0^{d/2-1} 
\end{equation} 
for any node $x$. Moreover, thanks to statistical homogeneity, it is clear
that on a given tree $w=(2u)^{-1}$, with $u$ a specific instance of a
random variable distributed according to $F(u,\infty)$. Since $F(u,\infty)$
is not $\delta-$like (see previous section), we have in general $w\ne {\bar
  w}$, that is local and average spectral weights differ in general, unlike
the spectral dimension. One might wonder then about the auto--averaging
property, since $w$ does not fluctuate over the tree. Of course the
resolution of this apparent contradiction is in the order of limits:
averaging over the infinite graph and small $\mu_0$ asymptotics of
$C_{xx}(\mu_0)$ are operations that cannot be commuted in general.

These last remarks suggests a further development in the explicit
calculation of the universal limit function $F(u,\infty)$. Another
interesting direction of research is in the construction of modified
tree--generating algorithms such that scaling is preserved with $\dc<2$, in
order to check the general formula (\ref{eq:general}).   

\begin{acknowledgments}
We thank R. Burioni and D. Cassi for many valuable comments and suggestions.
\end{acknowledgments}

\appendix

\section{Asymptotic behaviour of moments}
\label{app}

First of all let us write the explicit form of $\F_n$
\begin{equation*}
  \F_n(V_1,V_2,\ldots,V_n) = V_n + 
  \sum_{k=2}^n \sum_{n_1=1}^{n-1}\ldots\sum_{n_k=1}^{n-1}
  \delta\Big(n-\sum_{j=1}^k n_i\Big) \prod_{j=1}^k V_{n_j}
\end{equation*}
In the case without the non--extinction condition we will prove that
\begin{equation}\label{eq:pvn}
  \avgn{\prod_{j=1}^{k} V_{n_j}}{r} \sim 
  \alpha^{\sum_j n_j -1} \, r^{3 \sum_j n_j - k -1}
\end{equation}
First of all let us prove this equation when all $n_j=1$; the case
$k=1$ is known from equation (\ref{eq:v1n}) while for arbitrary $k$ is
proved by induction; in fact if 
\begin{equation}
  \label{eq:v1pn}
  \avgn{V_1^j}{r} \sim \alpha^{j-1} r^{2j-1}
\end{equation}
is assumed for $j\le k$ we have 
\begin{equation} \label{eq:v1prn}
  \begin{split}
    \avgn{V_1^{k+1}}{r+1} 
    & = \sum_z f_z \sum_{V_{1,1},V_{1,2},\ldots,V_{1,z-1}} 
    \left( 1+\sum_{i=1}^{z-1} V_{1,i} \right)^{k+1} 
    \prod_{i=1}^{z-1} \Pn_r(V_{1,i}) \\
    & = \sum_z f_z \sum_{V_{1,1},V_{1,2},\ldots,V_{1,z-1}} 
    \left( \left(\sum V_{1,i}\right)^{k+1} + k \left(\sum V_{1,i}\right)^k +
    \ldots + 1 \right) 
    \prod_{i=1}^{z-1} \Pn_r(V_{1,i}) \\
    & = \avgn{V_1^{k+1}}{r} 
    + \alpha \sum_{j=1}^k \avgn{V_1^j}{r} \avgn{V_1^{k+1-j}}{r} \\
    & \qquad + g'''(1) \sum_{j_1,j_2,j_3=1}^k \avgn{V_1^{j_1}}{r} 
    \avgn{V_1^{j_2}}{r} \avgn{V_1^{j_3}}{r} \delta(k+1-j_1-j_2-j_3) + \ldots \\
    & \qquad + k \avgn{V_1^k}{r} 
    + \alpha \sum_{j=1}^k \avgn{V_1^j}{r} \avgn{V_1^{k-j}}{r} + 
    \text{lower order terms} \\
    & \sim \avgn{V_1^{k+1}}{r} + \alpha^k r^{2k} 
  \end{split}
\end{equation}
where all addends which consist of a product of three or more averages
and the ones that come from lower powers of $\left(\sum
  V_{1,i}\right)$ are negligible in the limit $r\to\infty$ thanks to
equation (\ref{eq:v1pn}), so that
\begin{equation*}
  \avgn{V_1^{k+1}}{r} \sim \alpha^{(k+1)-1} \, r^{2(k+1)-1}
\end{equation*}
extending equation (\ref{eq:v1pn}) from $k$ to $k+1$.

Now, let us prove equation (\ref{eq:pvn}) by induction; in this case
the induction is on the quantity $N=3 \sum_{j=1}^k n_j - k$. Let us
denote with $\bar n$ the greatest among $n_1,n_2,\cdots,n_k$; we have
\begin{equation*}
  \begin{split}
    \avgn{\prod_{j=1}^k V_{n_j}}{r+1} & = \sum_z f_z
    \sum_{\substack{V_{1,1} \cdots V_{1,z-1} \\
        V_{2,1} \cdots V_{2,z-1} \\
        \cdots \\
        V_{\bar n,1} \cdots V_{\bar n,z-1}}} \prod_{j=1}^k
    \Big( \delta_{n_j,1} +\sum_{i=1}^{z-1} \F_{n_j}(V_{1,i},\cdots V_{n_j,i})
    \Big) \prod_{i=1}^{z-1} \Pn_r(V_{1,i},\cdots V_{\bar n,i}) \\
    & = \avgn{\prod_{j=1}^k V_{n_j}}{r} + \sum_{j=1}^k
    \sum_{l=1}^{n_j-1} \avgn{V_l V_{n_j-l}
      \prod_{j' \ne j} V_{n_{j'}}}{r} + \\
    & \qquad \alpha \sum_S \avgn{\prod_{j\in S} V_{n_j}}{r}
    \avgn{\prod_{j \notin S} V_{n_j}}{r} + \text{lower order terms} \\
    & \sim \avgn{\prod_{j=1}^k V_{n_j}}{r} + \alpha^{\sum_j n_j -1} \, 
    r^{3\sum_j n_j -k -2}
  \end{split}
\end{equation*}
where $S$ are all the possible ordered subsets of $\{n_1,n_2,
\ldots,n_k\}$.  Now equation (\ref{eq:pvn}) follows but a few remarks
are needed. First, all terms in the right hand side involve averages
whose value of N is smaller than the one in the left hand side. Then
all the terms that are not explicitly written are proportional to
lower power of $r$: some of them in fact consist in the product of
three or more averages (each of them carries a ``$-1$'' in the
exponent), some other are expressions with more than $k+1$ factors
$V_n$, and finally some others come from products involving the
$\delta_{n_j,1}$.

We turn now to the probability $\Pt(V_1,\ldots,V_n)$ with non--extinction
condition for which the recursions reads
\begin{equation*}
  \Pt_{r+1}(V_1,\ldots,V_n) = \sum_z \tilde f_z 
  \!\!\!\!\!\sum_{\substack{V_{1,1} \cdots V_{1,z-1} \\ 
      V_{2,1} \cdots V_{2,z-1} \\
      \cdots \\
      V_{n,1} \cdots V_{n,z-1}}}\!\!\!\!\!
  \Pt_r(V_{1,1},\ldots,V_{n,1})
  \prod_{i=2}^{z-1} \Pn_r(V_{1,i},\ldots,V_{n,i}) 
  \prod_{k=1}^n 
  \delta \Big(V_k - \delta_{n,1} - \sum_{j=1}^{z-1} 
  \F_k(V_{1,j},\ldots,V_{k,j}) \Big) 
\end{equation*}
We will prove that
\begin{equation} \label{eq:pvc}
  \avgt{\prod_{j=1}^{k} V_{n_j}}{r} \sim \alpha^{\sum_j n_j} \, 
  r^{3 \sum_j n_j - k}
\end{equation}
Some special cases, including $k=n_1=1$ have been already shown in
section \ref{scaling}, while the general case is proved by induction.
In fact the recurrence for $\avgt{\prod_{j=1}^{k}
  V_{n_j}}{r}$ reads (writing only leading order terms, as before)
\begin{equation*}
  \begin{split}
    \avgt{\prod_{j=1}^k V_{n_j}}{r+1} & = \sum_z (z-1) f_z
    \sum_{\substack{V_{1,1} \cdots V_{1,z-1} \\
        V_{2,1} \cdots V_{2,z-1} \\
        \cdots \\
        V_{\bar n,1} \cdots V_{\bar n,z-1}}} \prod_{j=1}^k
    \Big( \delta_{n_j,1} + \sum_{i=1}^{z-1} \F_{n_j}(V_{1,i},\cdots V_{n_j,i})
    \Big) \\
    & \qquad \Pt_r(V_{1,1},\cdots V_{\bar n,1}) 
    \prod_{i=2}^{z-1} \Pn_r(V_{1,i},\cdots V_{\bar n,i}) \\
    & = \avgt{\prod_{j=1}^k V_{n_j}}{r} + 
    \sum_{j=1}^k \sum_{l=1}^{n_j-1} \avgt{V_l V_{n_j-l}
      \prod_{j' \ne j} V_{n_{j'}}}{r} +
    \alpha \avgn{\prod_{j=1}^k V_{n_j}}{r} + \\
    & \qquad \alpha \sum_S \avgt{\prod_{j\in S} V_{n_j}}{r}
    \avgn{\prod_{j \notin S} V_{n_j}}{r} + \text{lower order terms} \\
    & \sim \avgt{\prod_{j=1}^k V_{n_j}}{r} + \alpha^{\sum_j n_j} \, 
    r^{3\sum_j n_j -k -1}
  \end{split}
\end{equation*}
with the same remarks as in the previous case.

Finally the recursion for the probability $\Pc(V_1,\ldots,V_n)$ reads
\begin{equation*}
  \Pc_{r+1}(V_1,\ldots,V_n) = \sum_z f_z 
  \!\!\!\!\!\sum_{\substack{V_{1,1} \cdots V_{1,z} \\ 
      V_{2,1} \cdots V_{2,z} \\
      \cdots \\
      V_{n,1} \cdots V_{n,z}}}\!\!\!\!\!
  \Pt_r(V_{1,1},\ldots,V_{n,1})
  \prod_{i=2}^z \Pn_r(V_{1,i},\ldots,V_{n,i})
  \prod_{k=1}^n 
  \delta \Big( V_k - \delta_{n,1} - \sum_{j=1}^z \F_k(V_{1,j},\ldots,V_{k,j})
  \Big) 
\end{equation*}
and, following the same steps as before,
\begin{equation*}
  \begin{split}
    \avgc{\prod_{j=1}^k V_{n_j}}{r+1} & = \sum_z f_z
    \sum_{\substack{V_{1,1} \cdots V_{1,z} \\
        V_{2,1} \cdots V_{2,z} \\
        \cdots \\
        V_{\bar n,1} \cdots V_{\bar n,z}}} \prod_{j=1}^k
    \Big( \delta_{n_j,1} + \sum_{i=1}^z \F_{n_j}(V_{1,i},\cdots V_{n_j,i})
    \Big) \\
    & \qquad \Pt_r(V_{1,1},\cdots V_{\bar n,1}) 
    \prod_{i=2}^z \Pn_r(V_{1,i},\cdots V_{\bar n,i}) \\
    & = \avgt{\prod_{j=1}^k V_{n_j}}{r} + \text{lower order terms} \\
    & \sim \alpha^{\sum_j n_j} \, r^{3 \sum_j n_j - k}
  \end{split}
\end{equation*}

\end{document}